\documentclass[12pt,preprint]{aastex}

\newcommand{\kms}{$\mathrm {km\,s}^{-1}$}
\newcommand{\mdot}{$\mathrm {M_{\odot}\,yr}^{-1}$}

\shorttitle{Mass loss history of Mira}
\shortauthors{Wareing et al.}

\begin{document}

\title{It's a wonderful tail: the mass loss history of Mira}

\author{C. J. Wareing\altaffilmark{1,2}, A. A. Zijlstra\altaffilmark{1}, 
T. J. O'Brien\altaffilmark{1} and M. Seibert\altaffilmark{3}}

\altaffiltext{1}{Jodrell Bank Centre for Astrophysics, Turing Building, 
University of Manchester, Oxford Road, Manchester, M13 9PL, UK; 
a.zijlstra@manchester.ac.uk,tim.obrien@manchester.ac.uk}
\altaffiltext{2}{Department of Applied Mathematics, University of Leeds, 
Leeds, LS2 9JT, UK; cjw@maths.leeds.ac.uk}
\altaffiltext{3}{Observatories of the Carnegie Institution of Washington, 
813 Santa Barbara Street, Pasadena, California 91101, USA; mseibert@ociw.edu}

\begin{abstract}

Recent observations of the Mira AB binary system have revealed a
surrounding arc-like structure and a stream of material stretching 2
degrees away in opposition to the arc. The alignment of the proper
motion vector and the arc-like structure shows the structures to be a
bow shock and accompanying tail. We have successfully hydrodynamically
modelled the bow shock and tail as the interaction between the
asymptotic giant branch (AGB) wind launched from Mira A and the surrounding
interstellar medium. Our simulations show that the wake behind the bow
shock is turbulent: this forms periodic density variations in the tail
similar to those observed. We investigate the possiblity of mass-loss
variations, but find that these have limited effect on the tail
structure. The tail is estimated to be approximately 450\,000 years
old, and is moving with a velocity close to that of Mira itself.  We
suggest that the duration of the high mass-loss phase on the AGB may
have been underestimated. Finally, both the tail curvature and the
rebrightening at large distance can be qualitatively understood if
Mira recently entered the Local Bubble. This is estimated to have occured
17\,pc downstream from its current location.
\end{abstract}


\section{Introduction}

In the Mira binary system, Mira A is an evolved star which is 
undergoing a period of enhanced mass-loss as it moves along the asymptotic
giant branch (AGB) on route to becoming a white dwarf. The companion star, 
previously classified as a white dwarf, now has a less clear classification
\citep{karovska05,ireland07} but is less luminous and any stellar outflow is 
comparably insignificant to Mira A in terms of mass-flux and energetics.

New UV observations of the Mira system \citep{martin07} have revealed
a comet-like tail extending 2 degrees away to the North and an
arc-like structure in the South.  \cite{martin07} postulated that
these features are a bow shock and a ram-pressure-stripped tail
stretching away in opposition to the bow shock caused by motion
through the interstellar medium (ISM). Their postulation is consistent
with Mira's proper motion \citep{turon93} of 225.8 milli-arcseonds per
year in the direction 187.1 degrees East of North (corrected for solar
motion). Further, they note that at the revised Hipparcos-based
distance \citep{knapp03} of 107 pc, the large space velocity of 130
\kms, calculated from the proper motion and the radial velocity
\citep{evans67} of 63 \kms, is further consistent with the bow shock
structure.

In this paper, we present hydrodynamical modelling of the Mira system
and discuss the implications with respect to postulation of
\cite{martin07}.  We are aiming to fit the position of the bow shock
ahead of the central star, the width across the central star, the
undulating density profile along the tail, the length of the tail 
and the ring-like structure one
third the way down the tail.  We show the GALEX observation in Figure
1.

\section{Simulations}

We have used a two-wind model consisting of a slow, dense AGB wind
ejected from the position of the mass-losing star and a second wind
representing the motion through the ISM. This model is the same as
that used to successfully model the circumstellar structures around
the evolved star R Hya \citep{wareing06b} and the planetary nebula Sh
2-188 \citep{wareing06,wareing07b}. We have employed the same
hydrodynamical scheme to simulate this model, with a numerical domain
of $800 \times 200 \times 200$ cells and the central star placed at
cell coordinates ($x$,$y$,$z$) = (50,100,100). Each cell is a regular
cube $6.25 \times 10^{-3}$ pc on a side giving a physical domain of $5
\times 1.25 \times 1.25$ pc$^3$. The scheme is second order accurate
and based on a Godunov-type method developed by \cite{falle91} using a
Riemann solver due to \cite{vanleer79}. The scheme has recently been
parallelised and extensively tested using standard computational fluid
dynamics tests and astrophysical tests \citep{wareing05} and includes
the effect of cooling above 10$^4$ K via cooling curves calculated by
\cite{raymond76}. Mass-loss is effected by means of artificially
resetting the hydrodynamical variables at the start of every timestep
in a volume-weighted spherical region of radius 5 3/4 cells centred on
the position of the central star.

In our $5 \times 10^5$ yr simulation, we have used a mass-loss rate in
the AGB wind of $3 \times 10^{-7}$ \mdot\ and a velocity of 5
\kms. These are based upon observations of the CO lines
\citep{ryde00}. An unphysical temperature of $10^4$ K was used for the
temperature of the wind as the cooling curves extend no further but
this does not affect the overall result \citep{wareing06b}.  For the
parameters of the ISM, we have considered a ram-pressure balance
between the AGB wind and the ISM. In the direction of motion, the
inner rim of the bow shock is 3.2 arc-minutes South of the system. At
D = 107 pc, this corresponds to a distance of $3.1 \times 10^{17}$ cm,
implying a local ISM density of n$_H$ = 0.03 cm$^{-3}$. The ISM density
depends directly on AGB wind mass-loss rate and velocity.

The simulation begins at the onset of mass-loss and is performed in
the frame of reference of the star.  The wind from the star drives a
shock into the ISM, which forms into a bow shock upstream of the
star. Such bow shocks have been simulated in other cases of winds
interacting with injected flows
\citep{villaver03,pittard05}. Eventually, the simulation shows the bow
shock reaches a maximum distance ahead of the star which can be
understood in terms of a ram pressure balance between the stellar wind
and the ISM. Strong shock theory predicts the temperature of the
shocked material at the head of the bow shock: T $\sim (3/16)\
m\,v^2/k$. Our simulation is consistent with
this. Ram-pressure-stripped material from the head of the bow shock
forms a tail behind the nebula and as material moves down the tail, it
deccelerates, cools and mixes with ISM material.

\section{Results}

\begin{figure}
\begin{center}
\epsscale{0.8}
\plotone{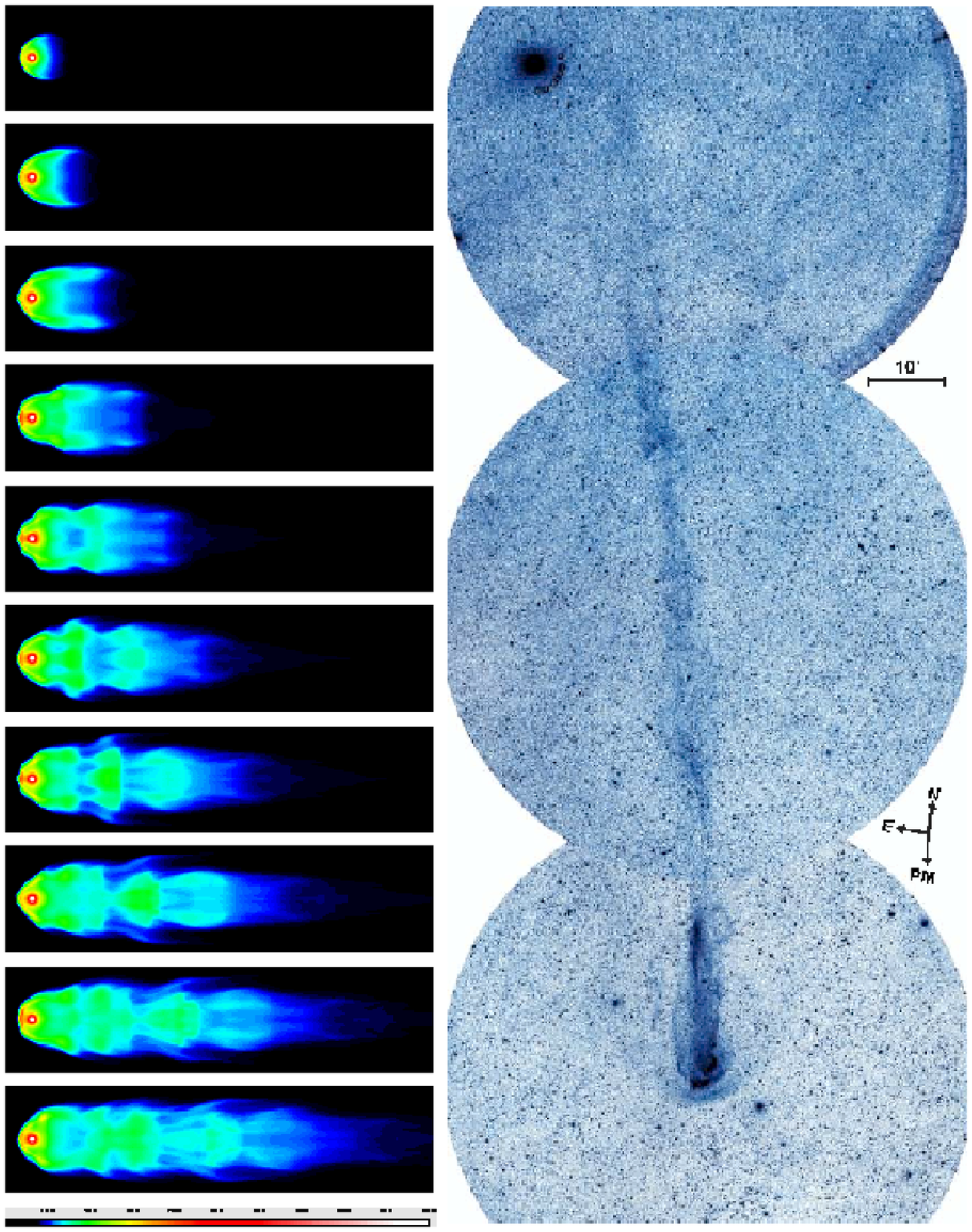}
\caption{In the left column we show snapshots of the density datacube 
collapsed along a line of sight. The top panel is 50,000 years into 
the AGB phase of evolution, with each following panel 50,000 years 
further on. Each panel is 1.25\,pc by 5\,pc. 
In the right column we show the mosaic'd UV images of the Mira AB
tail and bow shock. This figure is an adaptation of figure 1(a) from
\cite{martin07}. Please consult that reference for full details.
The bright star close to the termination point of the tail is HR 691.} 
\label{datacube}
\end{center}
\end{figure}

In the left column of Figure \ref{datacube} we show images of the
density datacube collapsed along a line of sight perpendicular to the
direction of motion and parallel to the y-axis of the simulation
domain, at 50,000 year intervals through the simulation. In the right
column of Figure \ref{datacube} we show an adaptation of figure 1(a)
from \cite{martin07} for comparison. The simulation best fits the
observational characteristics of the bow shock and tail approximately
450,000 years into its AGB evolution; in particular, this reproduces
the distance to the bowshock, the total width of the bowshock
(0.55\,pc across the position of the star) and the length of the tail
($\sim 4$ pc). Our choice of ISM density correctly reproduces the
position and width of the bow shock.  The temperature of the ISM is
less constrained as it has little effect on the ram pressure. The
narrowness of the tail suggests that the pressure in the ISM dominates
over the AGB wind confining it behind the star. This is
understandable, as Mira has a high velocity through the ISM combined
with a low mass-loss rate and outflow velocity for an AGB star.

The derived ISM density, 0.03\,cm$^{-3}$ is lower than derived by
\cite{martin07} who find 0.8\,cm$^{-3}$; we cannot reproduce the
location of the bow shock with such a high ISM density. The density is
comparable to that of the Local Bubble, a low-density region of about
100\,pc across in which the Sun is also located. An approximate 3-d
map of the Local Bubble is presented by \cite{lallement03}: on their
maps Mira would be located close to the edge but inside of the
higher density shell.

Our simulations imply the tail is far older than it appears from its
spatial extent. \cite{martin07} derive an age of $3 \times 10^4$\,yr,
based on the space velocity of Mira, assuming the tail is stationary
with respect to the ISM. However, we find this age to be a lower
limit. The material in the tail does not instantaneously decelerate to
the ISM velocity.  Instead, the gas being shed from the bow shock into
the tail travels only marginally slower than Mira itself, at a
relative velocity lag of 10-15 \kms. This lag increases as material
moves down the tail, which can be thought of as a decceleration with
time. Comparing this to our simulation, the tail's two degree extent
on the sky corresponds to an age of $4.5 \times 10^5$\,yr. This can
also be seen directly from Figure \ref{datacube} which shows the tail
has only grown to a length comparable to the observations after $\sim
4.5 \times 10^5$ years of evolution. Thus, to reproduce a 4 pc tail in
the simulation requires 450,000 years of mass-loss history. It is
possible that in reality, deceleration in the tail is faster than the
simulation indicates. This depends amongst others things on the
friction between the tail and the ISM. The viscosity in the simulation
is a fixed but artificial parameter, the value of which is chosen to
avoid the Quirk effect \citep{quirk94} which can cause accuracy issues
near the axes of the simulation.  A higher viscosity may yield a
narrower width of the tail. Measurements of the velocity of the tail
gas would help constrain this parameter.  This may also reduce the age
of the tail, which would be more comparable with the emission
mechanisms suggested by \cite{martin07}.  We therefore suggest that
the true age of the tail is somewhere between 30,000 and 450,000
years. For the rest of this paper, we use the latter age.

\begin{figure}
\begin{center}
\epsscale{0.4}
\plotone{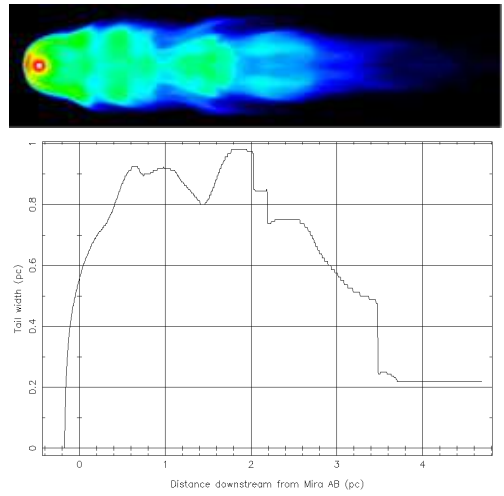}
\caption{The width of the tail. A collapsed density datacube image
is reproduced in the top panel at a point 450,000 years into the
AGB phase of evolution. Shown in the lower panel is a graph
of the width of the tail versus the position along its length.}
\label{width}
\end{center}
\end{figure}

In Figure \ref{width}, we show the variation of the tail width along
the length of the tail. The tail in our simulation is somewhat wider
than the observed width, but does reproduce similar periodic structure
along its length. The observed tail becomes more collimated with
increasing distance from the bow shock. We do not reproduce this
effect. This may be another consequence of friction between the tail
and the ISM.  \cite{martin07} has three possible explanations of the
variation along the length of the tail: mass-loss variations, 
turbulence in the ISM and variations in the ISM density. Our
simulation shows that undulations along the tail arise even when 
the mass loss and ISM density are assumed constant with time. 

We have also explored the effects of time-variable mass-loss rates.
The mass loss is expected to vary over the thermal pulse cycle
\citep{vassiliadis93}: a long phase of fairly constant mass loss
during the hydrogen burning phase, with a short spike during the
helium flash followed by much lower rate during the phase of quiescent
helium burning. As a guide, the H-burning lasts $\sim 10^5$\,yr, the
helium flash $10^3$\,yr and the quiescent helium burning $10^4$\,yr.
We assume that the ISM density is constant over the length of the tail
(4\,pc) and vary the mass-loss rate by a factor of three increase
during the Helium flash phase and a factor of three decrease during
quiescent helium burning.

\begin{figure}
\begin{center}
\epsscale{0.4}
\plotone{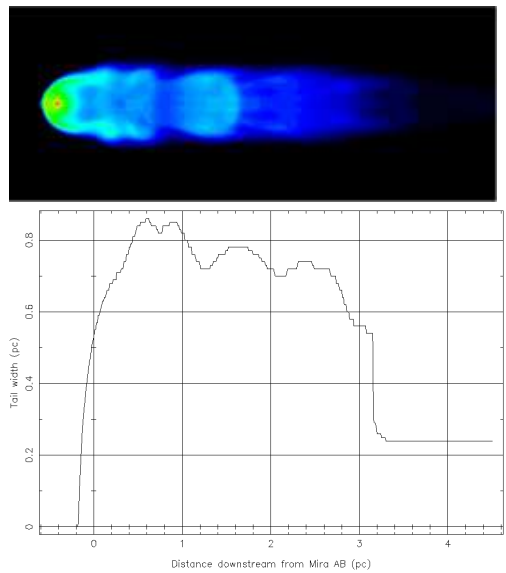}
\caption{Results of the second simulation with varying mass loss. Shown 
in the top panel is a collapsed density datacube image along a line of
sight parallel to the y-axis of the simulation and perpendicular 
to the direction of motion of the star at a stage 450,000 years into the
AGB phase of evolution. The image is 2 pc by 5 pc. In the lower panel is 
shown a graph of the width of the tail versus position along its length.}
\label{masslossvar}
\end{center}
\end{figure}

We show the results of this second simulation in Figure
\ref{masslossvar}.  Variation in the mass-loss rate has not greatly
affected the characteristics of the bow shock and tail.  Throughout
the second simulation, the bow shock is in the same position and has
the same width across the central star as it has in the first
simulation. Structures in the tail which were a result of turbulence
in the first simulation are in the same position, suggesting that
structures observed in Mira's tail are an effect of turbulence rather
than of variations in the mass-loss rate. The tail is slightly
narrower as can be seen from the lower panel of Figure
\ref{masslossvar}. This is in better agreement with the observations,
although still not as narrow as observed, possibly due to reasons
discussed previously.

\section{Discussion}

\subsection{Mass loss history}

The age of the tail has implications for the evolution of Mira. At its
current mass-loss rate, over $4.5 \times 10^5$\,yr Mira A will have lost
0.15\,M$_\odot$. A younger tail would proportionately reduce this value.
This is a signicant change in stellar mass,
sufficient to change the stellar radius and pulsation period. The
models of \cite{vassiliadis93} do not show any phase of mass loss at the rate
shown by Mira lasting for such a long time. Higher rates are reached
during the final phase of AGB evolution, the so-called superwind, but
these are short-lasting. This suggests that the theoretical mass-loss
rates for stars on the early thermal-pulsing AGB are underestimated.

Mira is a known binary and there has been a claim that the binary is 
causing asymmetric mass loss from the Mira system \citep{josselin00}.
We do not see any effect from the binary companion
or binary motion on the tail and bowshock and reproduce the observed
structure without introducing an asymmetry into the AGB wind. This is 
not unexpected, as the size and time scales of the tail are very 
much larger than those of the binary system.

\subsection{Mira and the Local Bubble}

Over the previous 450,000 years, Mira has travelled approximately 60
pc. The slow deceleration from the system velocity of 130 \kms\
explains why all the material released over this time is
still within 4 pc of Mira. Mira's Galactic position is (l,b) =
(168,-58) \citep{perryman97}, about 90\,pc South of the Galactic
plane.  Mira is traveling away from the Galactic plane: $v_z =
-96\,\rm km\,s^{-1} $.  The outermost extent of the tail was ejected
while Mira was much closer to the plane.  The star is likely to have
experienced considerable changes in ISM density on its trajectory. Two
aspects of the tail may best trace this: the narrowness and
rebrightening of the tail starting about 40$^\prime$ from Mira, and
the bend in the tail starting at the same position.  Assuming the 10
\kms\ deceleration of the tail, this point corresponds to
material ejected some 130,000\,yr ago, 17\,pc downstream from Mira's
current location. 

The observed changes along the tail can be interpreted as an
indication for a significantly higher ISM density at this location, 17
pc downstream, compared to Mira's current environment. A higher ISM
density would cause the bow shock to be closer to the star, and the
tail to become narrower as a result. The swept-up mass would be higher,
causing the brightening of the tail. 

It is interesting to compare this with Mira's location in the Local
Bubble.  Using the images of \cite{lallement03} (we use their figure
4), Mira is located within but close to the edge of the Local Bubble.
The velocity vector, based on proper motion and radial velocity, is
$v_z\approx -90 \,\rm km\,s^{-1}$ and $v_x \approx -25 \,\rm
km\,s^{-1}$ where a correction was applied for the solar motion. With
this motion, Mira may have passed into the Local Bubble some 30\,pc
back. This should not be overinterpreted as there are significant
uncertainties in the extent of the Local Bubble, and the map used is
for $l=180 \deg$ while Mira is at $l=168 \deg$. However, this shows
that the brightening along the tail can conceivably be caused by Mira
passing through the neutral wall of the Local Bubble.

This is also the approximate point where the tail begins to show
curvature.  The difference between the proper motion vector and the
tail direction measures the deceleration of the tail: the
tail gas follows a slightly different orbit from Mira itself,
remaining slightly closer to the Galactic plane than Mira was at that
point in its orbit.  A full simulation of this is beyond the scope of
this paper, or the hydrodynamical code used. The curvature is
probably caused by a larger deceleration of the gas in the tail, beyond
this point of 17\,pc downstream. A larger deceleration is expected for
a higher ISM density, and so is consistent with Mira having recently moved out
of a denser environment. 

Following the entry of of Mira into the Local Bubble, the bow shock
would have taken $\sim 10^4$\,yr to re-establish itself (or to grow to
its current size), and the tail may have taken $2-5\times 10^4$\,yr to
re-establish. This would correspond to the gap in the tail between
25$^\prime$ and 35$^\prime$ downstream. An equally likely explanation
is that the gap is complete H{\sc ii} dissociation in this low-density 
part of the tail.

The observations reveal a ring-like structure west of the tail, about
20 arcmin downstream. \cite{wareing07} show that under some
conditions, instabilities at the head of the bow shock cause
von-Karman-like vortex shedding into the ram-pressure stripped
tail. For the case of Mira, the simulations show turbulence, but this
does not grow further into vortices. However, \cite{wareing07} also
postulated that variations in the ISM could seed or enhance bow shock
instabilities and this could lead to vortex shedding. The reformation of
the bow shock after the entry into the Local Bubble could have been the
trigger for the initial instability leading to its single vortex.

\subsection{One of a kind?}

Observationally, the tail of Mira is unique. However, all evolved AGB
stars show similar {\it or stronger} winds to Mira. This raises the
question whether Mira is a proto-type or an exception, regarding its
tail. 

The main observational difference between Mira and typical similar
stars is in the space velocity. This will have two effects: first, the
tail narrows at higher velocities; second, the viscosity heats the
surface layers of the tail, as the tail gas itself also moves at
appreciable velocities with respect to the ISM. We suggest that this
secondary heating has strong effects on the observability.  Slower
tails may not generate sufficient heating to produce observable UV
emission.

We therefore suggest that Mira-like tails may be common, albeit in
most cases not as narrow. However, they are not evident in current
emission line surveys. An emission-line study of lower excitation
lines may be worthwhile.

It is of interest to speculate on the future appearance of the planetary
nebula Mira will eventually form. As the planetary expands it will
strongly brighten when it reaches the bow shock. At this stage, its
appearance may well be similar to that of the one-sided appearance of
Sh\,2-188 \citep{wareing06}. The evolution of planetary nebula within
a non co-moving ISM is discussed in \cite{wareing07b}.

\section{Conclusion}

Our models for Mira's evolution suggest that the
tail traces half a million years of mass-loss history and ISM interaction.
An instability at the bow shock is the cause of the fluctuations seen in
the tail. Mass-loss variations of Mira A have much less effect. The
curvature in the tail is caused by the differential velocity between
Mira and its tail leading to different Galactic orbits. We attribute
a gap in the tail to Mira entering the local bubble, 17 pc downstream
from its current location. The time to reestablish the bow shock 
directly leads to the gap in the tail. A ring-like structure is attributed
to a vortex shed into the tail just before the bow shock reached its 
equilibrium position. The tail therefore traces not only the history of
Mira itself, but also the structure of the ISM along its path. It's
a wonderful prospect.

\acknowledgments

The numerical computations were carried out using the COBRA supercomputer 
at Jodrell Bank Observatory.


\clearpage

\clearpage

\clearpage

\end{document}